\documentclass[pdftex,twocolumn,epjc3]{svjour3}
\pdfoutput=1 
\newif\ifcountpages
\countpagesfalse

\RequirePackage[T1]{fontenc}
\RequirePackage{graphicx}
\RequirePackage{mathptmx}
\usepackage{caption}
\RequirePackage{flushend}
\RequirePackage[numbers,sort&compress]{natbib}

\usepackage{amssymb}
\usepackage{lineno}
\usepackage{chngpage}
\usepackage{atlasphysics}
\usepackage{mathrsfs}
\usepackage{extra_definitions}
\usepackage{verbatim}
\usepackage{rotating}

\usepackage{dblfloatfix}
\usepackage{fixltx2e}

\def\ttbar {\mbox {$t\bar{t}$}}
\def\wp {\mbox {$W'$}}
\def\wpr {\mbox {$W'_{R}$}}
\def\wpl {\mbox {$W'_{L}$}}
\def\lumi{\mbox{$20.3\;\ifb$}}
\def\kt{\mbox{$k_T$}}

\def\mwp{\ensuremath{m_{W'}}}

\def\limitleft{\mbox{1.68~\tev}}
\def\limitright{\mbox{1.76~\tev}}
\def\explimitleft{\mbox{1.63~\tev}}
\def\explimitright{\mbox{1.85~\tev}}
\newcommand\gSM{\ensuremath{g_{\rm{SM}}}}

\newcommand\titletext{\boldmath Search for $W' \rightarrow tb \rightarrow qqbb$ Decays in $pp$ Collisions at $\sqrt{s}$~=~8~\TeV\ with the ATLAS Detector\unboldmath}
\newcommand\abstracttext{A search for a massive \wp\ gauge boson decaying to a top quark and a bottom quark is performed with the ATLAS detector in $pp$ collisions at the LHC.
The dataset was taken at a centre-of-mass energy of $\sqrt{s} = 8\tev$ and corresponds to $20.3\;\ifb$ of integrated luminosity.
This analysis is done in the hadronic decay mode of the top quark, where novel jet substructure techniques are used to identify jets from high-momentum top quarks.
This allows for a search for high-mass \wp\ bosons in the range $1.5 - 3.0\tev$.
$b$-tagging is used to identify jets originating from $b$-quarks. 
The data are consistent with Standard Model background-only expectations, and upper limits at 95\% confidence level
are set on the $\wp \rightarrow tb$ cross section times branching
ratio ranging from $0.16\;\mathrm{pb}$ to $0.33\;\mathrm{pb}$ for left-handed \wp\ bosons,
and ranging from $0.10\;\mathrm{pb}$ to $0.21\;\mathrm{pb}$ for \wp\ bosons with purely right-handed couplings.
Upper limits at 95\% confidence level are set on the \wp-boson coupling to $tb$ as a function of the \wp\ mass using an effective field theory approach,
which is independent of details of particular models predicting a \wp\ boson.}

\RequirePackage[colorlinks,breaklinks]{hyperref}  
\hypersetup{linkcolor=blue,citecolor=blue,filecolor=black,urlcolor=blue} 

\usepackage{preprintcover}  
\PreprintCoverPaperTitle{\titletext}  
\PreprintIdNumber{CERN-PH-EP-2014-152}  
\PreprintCoverAbstract{\abstracttext}  
\PreprintJournalName{European Physical Journal C}  

\clearpage

\begin{document}
\sloppy
\title{\titletext}
\author{The ATLAS Collaboration}
\institute{}
\date{Received: 6 August 2014 / Accepted: 26 March 2015}
\maketitle
\begin{abstract}
\abstracttext
\keywords{Exotics, \wp\ boson, top-tagging, jet substructure}
\end{abstract}

\ifcountpages
\null\newpage
\null\newpage
\fi

\section{Introduction}
\label{sec:intro}
Several theories beyond the Standard Model (SM)~\cite{Glashow:1961tr,Weinberg:1967tq,Salam:1968rm} involve enhanced symmetries that introduce new charged vector currents carried by new heavy gauge bosons, usually called \wp\ bosons. 
For instance, Grand Unified Theories~\cite{Pati:1974yy,Georgi:1974sy,Georgi:1974my,Fritzsch:1974nn} extend fundamental symmetries of the SM, in which a massive
right-handed counterpart to the SM $W$ boson may occur.
\wp\ bosons can appear in phenomenological models involving extra space-time dimensions such as Kaluza-Klein excitations of the SM $W$ boson~\cite{Burdman:2006gy} or in technicolor models~\cite{Malkawi:1996fs}. 
Also Little Higgs theories~\cite{Perelstein:2005ka} predict several new particles, including a \wp\ boson.
In order to interpret a direct experimental search independently of the details of particular models predicting a \wp\ boson, it is advantageous to rely on an effective model describing the couplings of the \wp\ boson to fermions~\cite{Sullivan:2002jt}.

The search for a \wp\ boson decaying to a top quark and a $b$-quark ($\wp \rightarrow tb$)\footnote{
Decays of \wp$^{+}$ to a top quark and an anti-$b$ quark and of \wp$^{-}$ to an anti-top quark and a b-quark are equally taken into account. For simplicity, both decays are referred to as $\wp \to tb$ in this document.
} explores models potentially inaccessible to $\wp \rightarrow \ell\nu$ searches. 
Also, in the right-handed sector, it is assumed that there is no light right-handed neutrino to which a \wp\ boson could decay, and, hence, only hadronic
decays are allowed~\cite{Sullivan:2002jt,Duffty:2013aba}.
In some theories beyond the SM, new physics couples more strongly to the third generation than to the first and second~\cite{Malkawi:1996fs}. 
Searches for \wp\ bosons decaying to $tb$ have been performed at the Tevatron~\cite{D0wprime,CDFwprime,Aaltonen:2015xea} and at the LHC~\cite{CMSWprimePaper,Aad:2014xea}, in leptonic top-quark decay channels excluding a \wp\ boson with purely right-handed couplings (referred to as \wpr) with mass less than 2.13~\tev\ at 95\% Confidence Level (CL).

This document describes the first search for the $\wp \rightarrow tb $ process in the fully hadronic final state of the top-quark decay.
For high \wp\ masses, the final state signature consists of one high-momentum $b$-quark and another $b$-quark close to the two light-quarks from the $W$-boson decay.
The distinct signature of high-momentum top quarks is exploited to isolate the signal from the copious hadronic multijet background making use of novel jet substructure techniques to identify boosted hadronically decaying top quarks.
This allows for particularly good sensitivity at high \wp\ masses.
95\% CL exclusion limits are presented on the \wp-boson coupling as a function of the \wp\ mass in an effective model.

\section{The ATLAS Detector}
\label{sec:ATLAS}
Charged particles in the pseudorapidity\footnote{
  The ATLAS experiment uses a right-handed coordinate system with its origin at the nominal interaction point (IP) in the centre
  of the detector, and the $z$-axis along the beam line. The $x$-axis points from the IP to the centre of the LHC ring, and the
  $y$-axis points upwards. Cylindrical coordinates $(r, \phi)$ are used in the transverse plane, $\phi$ being the azimuthal angle
  around the beam line. Observables labelled ``transverse'' are projected into the $x$~-~$y$ plane. The pseudorapidity is defined
  in terms of the polar angle $\theta$ as $\eta=-\ln\tan \theta/2$. The transverse momentum is defined as $\pt=p\sin\theta=p/\cosh\eta$,
  and the transverse energy $E_T$ has an analogous definition.}
range \mbox{$|\eta| < 2.5$} are reconstructed with the inner detector (ID), which consists of several layers of semiconductor detectors (pixel and strip) and a straw-tube transition-radiation tracker, the latter extending to \mbox{$|\eta| < 2.0$}. 
The inner tracking system is immersed in a \mbox{2~$\mathrm{T}$} magnetic field provided by a superconducting solenoid.
The solenoid is surrouned by sampling calorimeters, which span the pseudorapidity range up to \mbox{$|\eta|$ = 4.9}.
High-granularity liquid-argon (LAr) electromagnetic calorimeters are present up to \mbox{$|\eta|$ = 3.2}.
Hadronic calorimeters with scintillating tiles as active material cover \mbox{$|\eta| <$ 1.74} while LAr technology is used for hadronic calorimetry from
\mbox{$|\eta|$ = 1.5} to \mbox{$|\eta|$ = 4.9}.
Outside the calorimeter system, air-core toroids provide a magnetic field for the muon spectrometer (MS). 
Three stations of precision drift tubes and cathode strip chambers provide a measurement of the muon track in the region of \mbox{$|\eta| < 2.7$}. 
Resistive-plate and thin-gap chambers provide muon triggering capability up to \mbox{$|\eta| < 2.4$}.

\section{Data and Monte-Carlo Simulation Samples}
\label{sec:DataandMC}

\subsection{Data Samples}
\label{sec:Data}
The data used for this analysis was collected in $pp$ collisions in 2012 at a centre-of-mass energy of $\sqrt{s} = 8\tev$. 
All candidate events must satisfy data-quality requirements that include being recorded during the LHC stable-beam periods and proper functioning of the detector and trigger subsystems. 
After the trigger and data-quality requirements, the amount of data used by this analysis corresponds to an integrated luminosity of \lumi\ with
an average number of interactions per bunch-crossing of 20.7.

\subsection{Signal Modelling}
\label{sec:SignalMC}
The right- and left-handed \wp\ boson (denoted as \wpr\ and \wpl, respectively) models
are implemented in {\sc MadGraph~5}~\cite{MadGraph5} using FeynRules~\cite{Alloul:2013bka,Degrande:2011ua},
which is used to generate events at leading-order (LO) in $\alpha_s$ through Drell-Yan like production. 
{\sc MadGraph} also simulates the decay of the top quark taking spin correlations into account. 
{\sc Pythia~8.165}~\cite{Pythia8} is used for parton showering and hadronisation. 
{\sc CTEQ6L1}~\cite{cteq6l} parton distribution functions (PDFs) are used for the event generation. 

The \wpr\ and \wpl\ cross sections times branching ratios to the $tb$ final state are obtained from next-to-leading order (NLO) QCD calculations~\cite{Sullivan:2002jt,Duffty:2012rf} and are shown for different \wp\ masses in Table~\ref{tab:cross_section}.
The mass of a possible right-handed neutrino is assumed to be larger than the mass of the \wpr\ boson, allowing only hadronic decays of the \wpr. 
In the case of a \wpl\ boson, leptonic decays are allowed.
Dedicated Monte-Carlo (MC) simulation samples with interference effects between \wpl\ and SM $W$ included have been used to estimate the change in the number of expected
signal events. In the high mass signal region the change in event yield is less than 1\% after kinematic requirements. Interference effects with the SM $s$-channel single-top quark process are ignored.
All simulated samples are normalised to these NLO calculations using NLO/LO $k$-factors ranging from 1.15 to 1.35 depending on the mass and the chirality of the \wp\ boson.
The models assume that the \wp-boson coupling strength to quarks is the same as for the SM $W$ boson: $g'_R = \gSM$ and $g'_L = 0$ ($g'_R = 0$ and $g'_L = \gSM$) for \wpr\ (\wpl) bosons, where \gSM\ is the SM $SU(2)$ coupling.

\begin{table}[t]
\begin{center}
\caption{\label{tab:cross_section} NLO cross sections times branching ratio to $tb$ for different \wp\ masses for the left-handed and for the right-handed
  model~\cite{Sullivan:2002jt,Duffty:2012rf}.}
\begin{tabular}{ccc}
\hline
Mass [\tev] & $\sigma \times {\rm{BR(}} \wpl \to tb {\rm{)}}$ & $\sigma \times {\rm{BR(}} \wpr \to tb {\rm{)}}$ \\
\hline
1.5 & 0.40 pb & 0.52 pb \\
2.0 & 0.067 pb & 0.086 pb \\
2.5 & 0.014 pb & 0.017 pb \\
3.0 & 0.0035 pb & 0.0039 pb \\
\hline
\end{tabular}  
\end{center}
\end{table}

\subsection{Background Samples}
\label{sec:BackgroundMC}

The background estimate in this analysis is derived from a fit to data.
However, an initial background estimate is introduced in Section~\ref{sec:hadbackground}, which uses a data-driven technique based on sideband regions for the
multijet process and MC simulation samples for top-quark pair production (\ttbar).
 For this purpose, \ttbar\ production is simulated using the {\sc Powheg-Box} generator~\cite{Powheg1,Powheg2} coupled to {\sc Pythia~6.426}~\cite{Sjostrand:2006za,Skands:2010ak} for parton showering and hadronisation.
This sample uses the {\sc CTEQ6L1} PDF set.
The \ttbar\ samples are normalised to the next-to-next-to-leading order (NNLO) calculations in $\alpha_s$ including resummation of next-to-next-to-leading
logarithmic soft gluon terms with {\sc top++2.0}~\cite{Cacciari:2011hy,Baernreuther:2012ws,Czakon:2012zr,Czakon:2012pz,Czakon:2013goa,Czakon:2011xx}: $\sigma_{\ttbar}= 253^{+14}_{-16}$~pb.
PDF and $\alpha_s$ uncertainties are calculated using the {\sc PDF4LHC} prescription~\cite{Botje:2011sn} with the {\sc MSTW2008} 68\% CL NNLO~\cite{Martin:2009iq,Martin:2009bu},
{\sc CT10} NNLO~\cite{Lai:2010vv,Gao:2013xoa} and {\sc NNPDF2.3}~\cite{Ball:2012cx} PDF sets, added in quadrature to the scale uncertainty. An uncertainty on the top-quark mass
of $ 1\gev$ is also considered.

For the optimisation of the \wp\ top-tagger (Section~\ref{sec:hadtoptag}), MC samples are generated with {\sc Pythia~8.160} using the {\sc AU2} tune~\cite{ATLAS:2012uec} and the {\sc CT10}~\cite{Lai:2010vv} PDF set.

After event generation, all signal and background MC samples are passed through a full simulation of the ATLAS detector~\cite{ATL-2010-005} based on {\sc GEANT4}~\cite{SAMPLES-G4} and then reconstructed using the same algorithms as for collision data. 
All MC processes are simulated with pile-up interactions included and re-weighted to match the conditions of the data sample.

\section{Physics Objects and Boosted Top Identification}
This analysis relies on the reconstruction and identification of jets.
Jets are built from energy depositions in the calorimeters with the anti-\kt\ algorithm~\cite{Salam:2007xv} using locally-calibrated topological clusters as inputs.
Jets are further calibrated using energy and $\eta$-dependent correction factors derived from simulation and with residual corrections from \textit{in-situ}
measurements~\cite{ATLAS-CONF-2013-004}.
Events with jets built from noisy calorimeter cells or non-collision backgrounds are removed~\cite{ATLAS-CONF-2010-038}.
In this analysis two radius parameters are used for jet reconstruction: a small-$R$ radius of 0.4 and a large-$R$ radius of 1.0.
Small-$R$ jets are required to have $\pt > 25\gev$\ and $|\eta| < 2.5$. 
To minimise the impact of energy depositions from pile-up interactions the large-$R$ jets are trimmed~\cite{Krohn:2009th}.
The trimming algorithm reconstructs jets using the $k_T$ jet algorithm with $R=0.3$ built out of the constituents of the original large-$R$ jet.
Constituent jets contributing less than 5\% of the large-$R$ jet \pt\ are removed.
The remaining energy depositions are used to calculate the jet kinematics and substructure properties. 
Large-$R$ jets are required to have $\pt > 350\gev$ and $|\eta| < 2.0$.

In order to identify small-$R$ jets which originate from $b$-quarks, this analysis uses a neural-network based $b$-tagging algorithm~\cite{ATLAS-CONF-2014-046}.
Different observables based on the long lifetime of $B$ hadrons are used as inputs and are able to discriminate between $b$-jets, $c$-jets and light-quark jets.

Events with reconstructed high-quality electrons~\cite{Aad:2014fxa}
or muons~\cite{Aad:2014zya} are vetoed in order to ensure orthogonality to analyses using the leptonic decay of the top quark~\cite{Aad:2014xea}.
Electrons and muons with transverse momenta above 30~\gev\ are considered for this veto.

\subsection{The \wp\ Top-tagger}
\label{sec:hadtoptag}
This analysis searches for \wp\ bosons in the high mass ($\mwp > 1.5\tev$) region, where the top quark and bottom quark have high transverse momentum. 
The average distance between the decay products of the top quark falls with increasing top-quark \pt, and their hadronic showers begin to overlap. 
This high-\pt\ topology, where the decay products of a massive particle can be captured in one single large-$R$ jet, is referred to as ``boosted''~\cite{Rehermann:2010vq,atlasboosttopconf,groomedCONF2011,groomedJetPileup2011}.

The discrimination of large-$R$ jets originating from hadronic top-quark decays from large-$R$ jets originating from other sources using calorimeter information is termed top-tagging.
The \wp\ top-tagging algorithm is a cut-based algorithm using different large-$R$ jet substructure properties developed to efficiently select large-$R$ jets
from \wp\ signal events over the dominant background from multijet production featuring light-quark, $b$-quark and gluon-initiated jets. 
The procedure uses three substructure variables: the one-to-two $\kt$-splitting scale $\sqrt{d_{12}}$~\cite{Butterworth:2002tt} and two ratios of $N$-subjettiness ($\tau_N$) variables~\cite{Thaler:2010tr,Thaler:2011gf} $\tau_{32} = \tau_3 / \tau_2$ and $\tau_{21} = \tau_2 / \tau_1$.

The splitting scale $\sqrt{d_{12}}$ distinguishes jets containing top-quark decays, which are relatively \pt-symmetric in the top-quark rest frame, from
\pt-asymmetric light jets.
It is calculated by reclustering the constituents of the large-$R$ jet using the \kt\ algorithm, where the reclustering procedure is stopped at the last merging step.
Since the \kt\ algorithm clusters the hardest objects last, the last clustering step corresponds to the merging of the two hardest subjets, and $\sqrt{d_{12}}$ is defined
as the corresponding scale:
\begin{equation*}
  \label{eq:splittingscale}
  \sqrt{d_{12}} = \text{min}(p_{\mathrm{T},1},p_{\mathrm{T},2}) \times \sqrt{\left( \Delta \eta_{12} \right)^2 + \left( \Delta \phi_{12} \right)^2} \; ,
\end{equation*}
\noindent
where $p_{\mathrm{T},1}$ and $p_{\mathrm{T},2}$ are the transverse momenta of the two remaining subjets, and $\Delta \eta_{12}$ and $\Delta \phi_{12}$ are the
distances in $\eta$ and $\phi$ between these two subjets.
For jets from hadronic top-quark decays the $\sqrt{d_{12}}$ distribution is expected to peak at approximately half the top-quark mass.
For jets initiated by light quarks, $b$-quarks and gluons, the $\sqrt{d_{12}}$ distribution is expected to peak near zero.

 \begin{figure}[!h]
  \centering
  \includegraphics[width=0.48\textwidth]{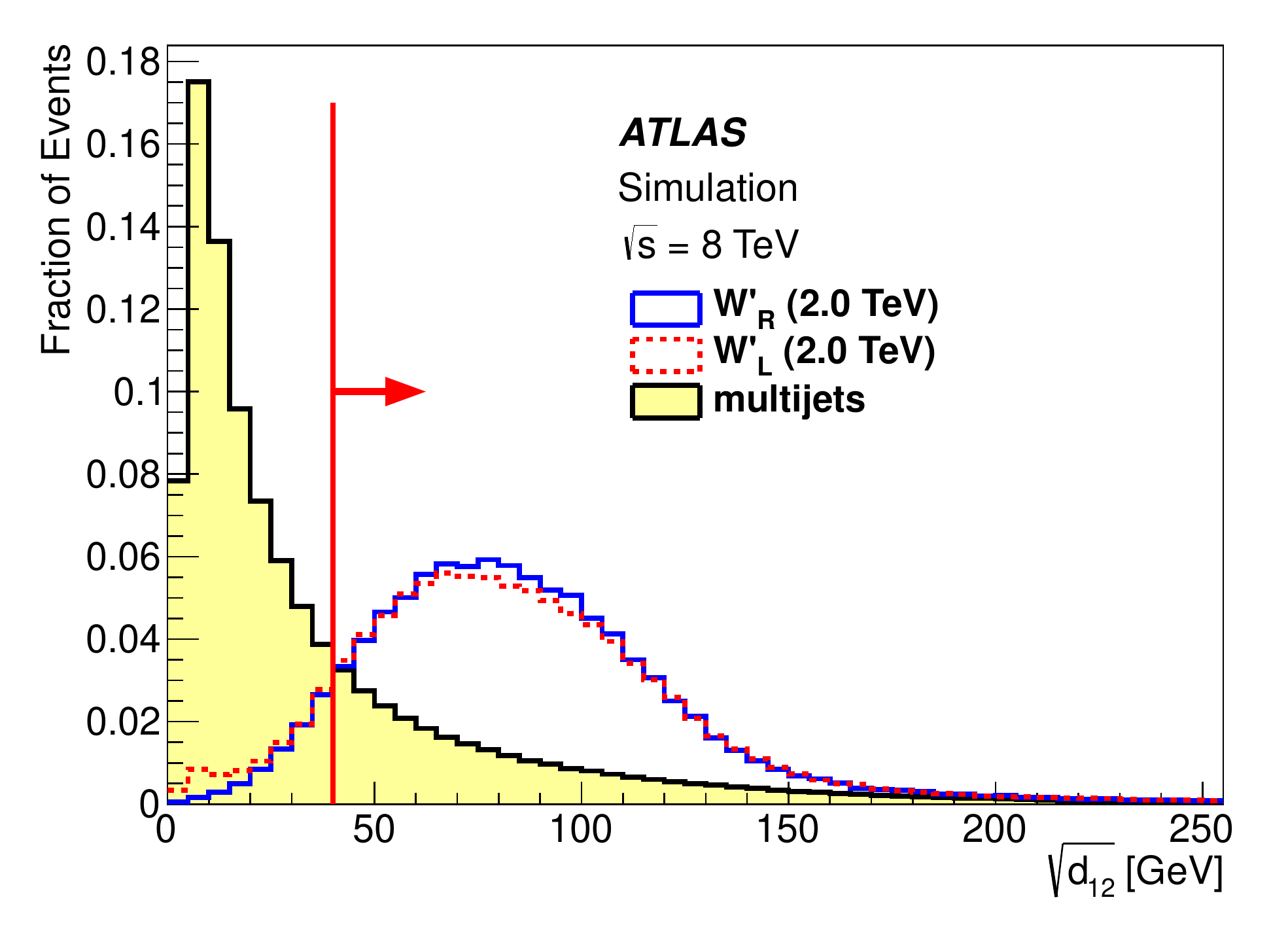}
  \includegraphics[width=0.48\textwidth]{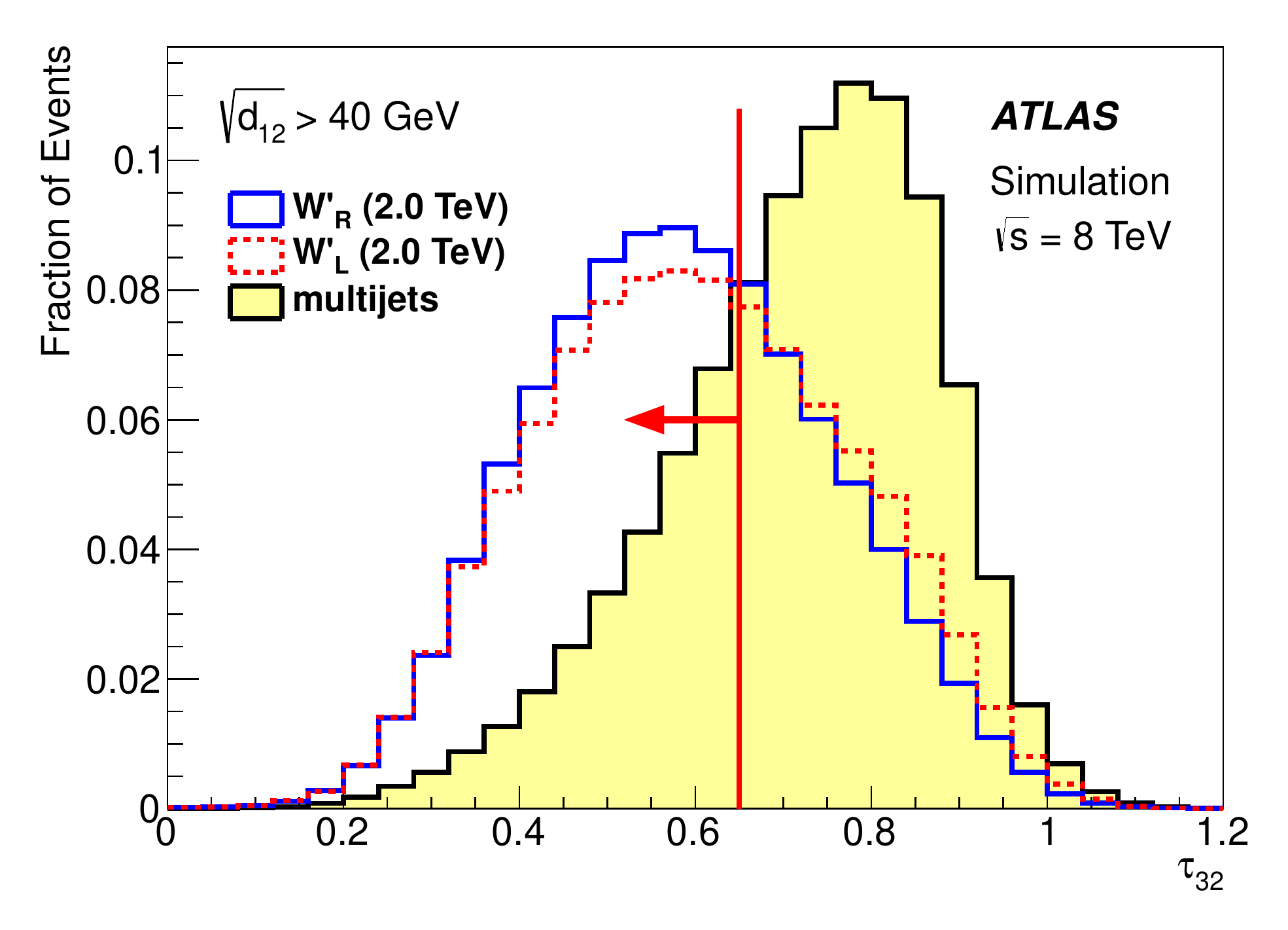}
  \includegraphics[width=0.48\textwidth]{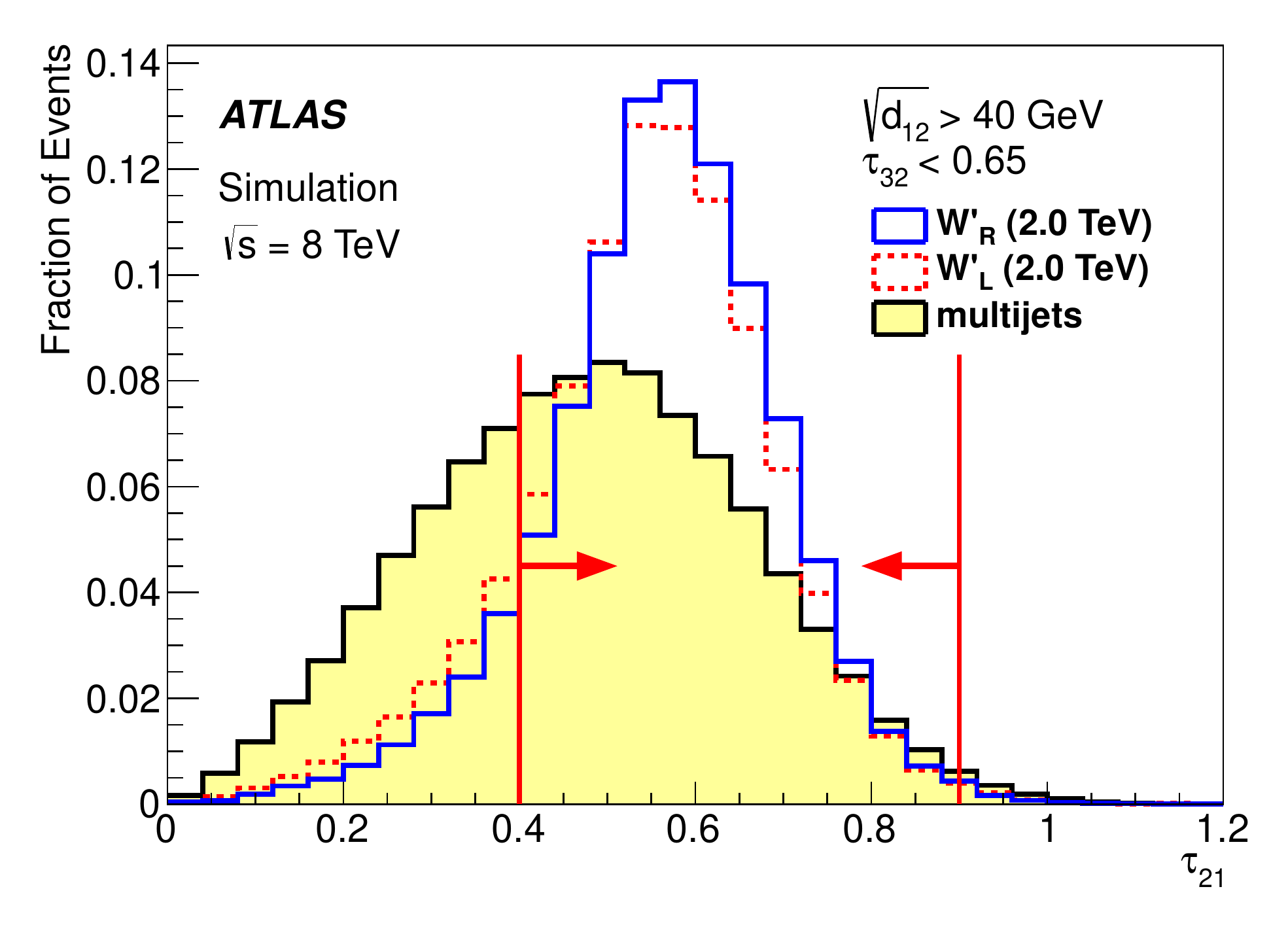}
  \caption{Distributions in simulated samples of the substructure observables used for the \wp\ top-tagger for trimmed large-$R$ jets originating from QCD-multijet production
    and originating from hadronic decays of top quarks from 2~\TeV\ \wpl\ and \wpr\ boson decays.
    The top figure shows the $\sqrt{d_{12}}$ distributions and the middle (bottom) figures show the $\tau_{32}$ ($\tau_{21}$) distribution after requiring
    $\sqrt{d_{12}} > 40\gev$ ($\sqrt{d_{12}} > 40\gev$ and $\tau_{32} < 0.65$).
    All distributions are normalised to unity.
    The arrows indicate the cut values used in the \wp\ top-tagger.
  }
  \label{fig:substructure}
\end{figure}

$N$-subjettiness is a measure of the compatibility of a large-$R$ jet with a given number of subjets.
The $\tau_N$ are calculated by reclustering the large-$R$ jet constituents with the \kt\ algorithm requiring exactly $N$ subjets to be found. The $\tau_N$ are then defined by:
\begin{equation*}
  \label{eg:nsubjettiness}
  \tau_N = \frac{1}{d_0}\sum_{k} p_{\mathrm{T}k} \times \text{min}(\delta R_{1k}, \dots ,\delta R_{Nk})~,
\end{equation*}
\noindent
with $d_0 = \sum_k p_{\mathrm{T}k} \times R$,
where the sum runs over all constituents of the jet, $p_{\mathrm{T}k}$ is the \pt\ of the $k^{\rm{th}}$ constituent, $R$ is the radius parameter of the original jet, and the variable $\delta R_{ik}$ is the distance in $\eta$-$\phi$ space from the $i^{\rm{th}}$ subjet to the $k^{\rm{th}}$ constituent. 
Ratios of the $\tau_N~(\tau_{ij}=\tau_{i}/\tau_{j})$ are then defined to discriminate if a jet is more $i$- or $j$-subjet-like. 
The $\tau_{ij}$ distributions peak closer to 0 for $i$-subjet-like jets and closer to 1 for $j$-subjet-like jets.

The optimisation procedure for the \wp\ top-tagger aims for an optimal compromise between the efficiency for jets originating from hadronically decaying top quarks
and the rejection of jets originating from QCD-multijet production.
First, an optimal requirement on $\sqrt{d_{12}}$ is applied and then, selection criteria on the $N$-subjettiness variables are determined.
The MC samples used are the 2~\tev\ $\wpl \rightarrow tb$ signal sample and a high-\pt\ QCD-multijet sample with a similar range in transverse momentum.
It has been checked that changing the order in which $\sqrt{d_{12}}$, $\tau_{32}$ and $\tau_{21}$ are optimised yields very similar results.

Figure~\ref{fig:substructure} shows distributions of $\sqrt{d_{12}}$ (top), $\tau_{32}$ with the $\sqrt{d_{12}}$ requirement applied (centre), and $\tau_{21}$ with both $\sqrt{d_{12}}$ and $\tau_{32}$ requirements applied (bottom) for jets originating from hadronically decaying top quarks in 2~\tev\ \wpl\ and \wpr\ MC simulations.
These are compared to the distributions for jets originating from light-quark, $b$-quark and gluon jets from QCD-multijet MC simulations.
The optimised top-tagging requirements are $\sqrt{d_{12}} > 40\gev$, $\tau_{32} < 0.65$ and $0.4 < \tau_{21} < 0.9$.
While $\tau_N$ is an infrared- and collinear-safe observable~\cite{Thaler:2010tr}, infrared-safety of $\tau_{32}$ is ensured by the requirements on $\tau_{21}$.
The selection efficiency for jets originating from hadronic top-quark decays is estimated in MC simulations to be larger than 50\% for jet \pt\ above 500~\gev, while
the probability to falsely tag a light-quark, $b$-quark or gluon jet is below 10\%~\cite{atlasboosttopconf}. For jet \pt\ below 800~\gev, where the sample size is sufficient, the top-tagging efficiency is cross-checked in data using single lepton $\ttbar$ events, and the top-tagging efficiency is found to be consistent between data and MC.

\section{Analysis}
\label{sec:analysis}
\subsection{Event Selection}
\label{sec:hadevent}
\begin{figure}[t]
  \centering
  \includegraphics[width=0.48\textwidth]{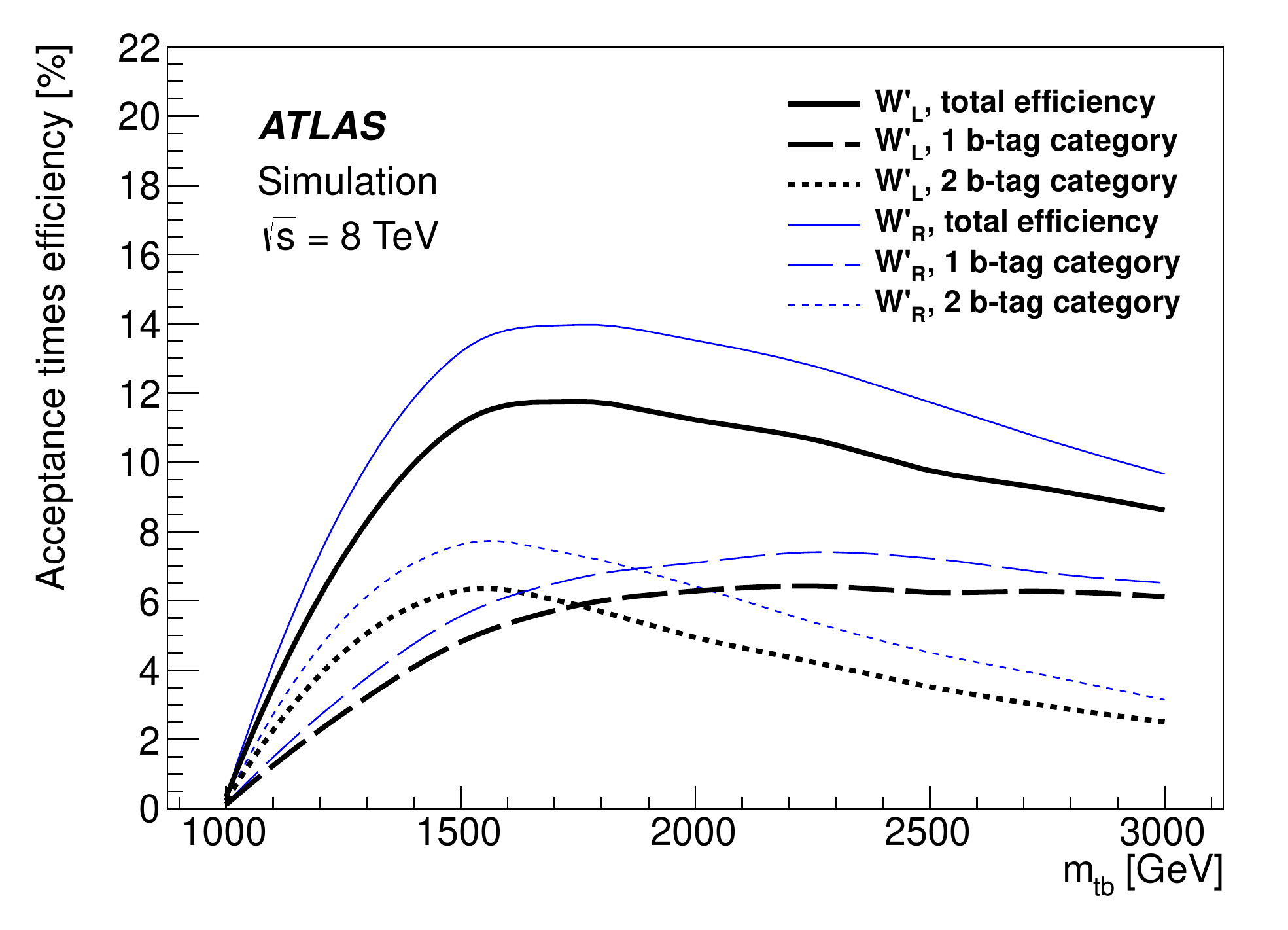}
  \caption{Selection acceptance times efficiency as a function of $tb$ invariant mass at truth level for left- and right-handed \wp\ MC. The total efficiency curves correspond to the sum of the efficiencies of the one $b$-tag and two $b$-tag categories.}
  \label{fig:sigeff}
\end{figure}

Candidate events are triggered by requiring the scalar sum of the $\et$ of the energy deposits in the calorimeters at trigger level to be at least 700~\gev. In order to perform the offline analysis in the fully efficient regime of this trigger, the scalar sum of the transverse momenta of reconstructed small-$R$ jets with $\pt > 25 \gev$ and $|\eta| < 2.5$ is required to be at least 850~\gev.
Candidate events must have at least one primary vertex with at least five tracks associated to it and have exactly one large-$R$ \wp\ top-tagged jet (top candidate) and one small-$R$ $b$-tagged jet ($b$-candidate) each with $\pt > 350~\gev$ and an angular separation
$\Delta R = \sqrt{\left(\Delta \eta\right)^2 + \left(\Delta \phi\right)^2}$ larger than 2.0 between the flight direction of the top candidate and the $b$-candidate.
The invariant mass of the dijet system must be at least 1.1~\tev\ in order to avoid turn-on effects from the kinematic selection.
The events are divided into two categories: the one $b$-tag category and the two $b$-tag category. 
For the two $b$-tag category, an additional $b$-tagged small-$R$ jet with $\pt > 25\gev$\ has to be present close to the top candidate by requiring $\Delta R$ between the small-$R$ $b$-jet and the top candidate to be less than the large-$R$ jet radius parameter 1.0. 
Figure~\ref{fig:sigeff} shows the acceptance times selection efficiency as a function of $tb$ invariant mass at truth level in the one and two $b$-tag categories and the total signal efficiency corresponding to their sum.  
The difference in the efficiencies observed in the \wpl\ and \wpr\ models originates from the different top-tagging efficiencies, which is due to the preferred flight
directions of the top-quark decay products in the top-quark rest frame for the two chiralities.

\subsection{Initial Background Estimate}
\label{sec:hadbackground}

An initial background estimate is performed to choose the fit function which is used to estimate the background from data.
It is also used to derive the associated systematic uncertainties on the background modelling.

Multijet events are the dominant background comprising 99\% (88\%) of the events in the one (two) $b$-tag category.
The estimate of the contribution of multijet events is based on a data-driven method which categorises events based on top- and $b$-tagging.
Requiring the high-\pt\ small-$R$ jet to fail the $b$-tagging requirement or the large-$R$ jet to fail the top-tagging requirement provides
three orthogonal control regions in each $b$-tag category that are dominated by multijet events.
These control regions are then used to make an estimate of the multijet contribution in the signal region, as defined by the event selection.

The other significant background is top-quark pairs, contributing 11\% in the two $b$-tag category as estimated using MC simulations.
Other backgrounds, such as single-top, $W$-boson$+$jets, and $Z$-boson$+$jets production are found to have a very small contribution.

Table~\ref{tab:final_eventyieldSM} reports the number of data and expected background events in the signal region for the one $b$-tag and two $b$-tag
categories.

\begin{table}[!h]
 \begin{center}
 \caption{\label{tab:final_eventyieldSM} Event yields in the signal region for SM processes using the initial background estimate compared to the yield observed in data.
The uncertainties quoted for multijet and for \ttbar\ production contain statistical as well as systematic uncertainties (Section~\ref{sec:systuncert}).
The contributions from other background sources (single-top, $W$-boson$+$jets and $Z$-boson$+$jets production) have only been estimated approximately,
but were found to be smaller than 0.4\% of the expectation for multijet production. An uncertainty of 100\% is quoted here in order to reflect the
approximations made.}
  \begin{tabular}{lcc}
  \hline
  Process & One $b$-tag & Two $b$-tag \\
  \hline
  Multijet & 16100 $\pm$ 800 & 2600 $\pm$ 300 \\
  Hadronic \ttbar\ & $130 \pm 30$ & $210 \pm 60$ \\
  Leptonic \ttbar\ & $60 \pm 20$ & $90 \pm 30$ \\
  Other & $60 \pm 60$ & $8 \pm 8$ \\
  \hline
  Total SM prediction & $16400 \pm 800$ & $2900 \pm 300$ \\
  \hline
  Data & 16601 & 2925 \\
  \hline
  \end{tabular}  
 \end{center}
\end{table}

\subsection{Statistical Analysis}
\label{sec:analsys}

An unbinned likelihood fit to the $m_{tb}$ distributions combining the one $b$-tag category and the two $b$-category is performed, where the range considered is 1.1~--~4~\tev.
The lower bound of $m_{tb}$ is due to turn-on effects of the $m_{tb}$ distribution originating in the kinematic selection.
This allows \wp\ signals of $m_{\wp} \geq 1.5\tev$ to be tested, because for lower values of $m_{\wp}$, the peak of the signal shape is only partially contained in the $m_{tb}$ range considered.
The upper bound of $m_{tb}$ is motivated by the low expected number of events in the two $b$-tag category at such high values.
Hence, \wp\ signals of $m_{\wp} \leq 3.0\tev$ can be tested, in order to constrain the background function parameters also for high $m_{tb}$ values.
A profile likelihood-based test statistic is used for the evaluation of $p$-values for observations as well as a $CLs$ test statistic for setting 95\% CL exclusion limits, where asymptotic formulas~\cite{asymlimits} are used. The expected and observed limits
are corrected for small differences observed using toy experiments.

\begin{figure}[b]
  \centering
  \includegraphics[width=0.48\textwidth]{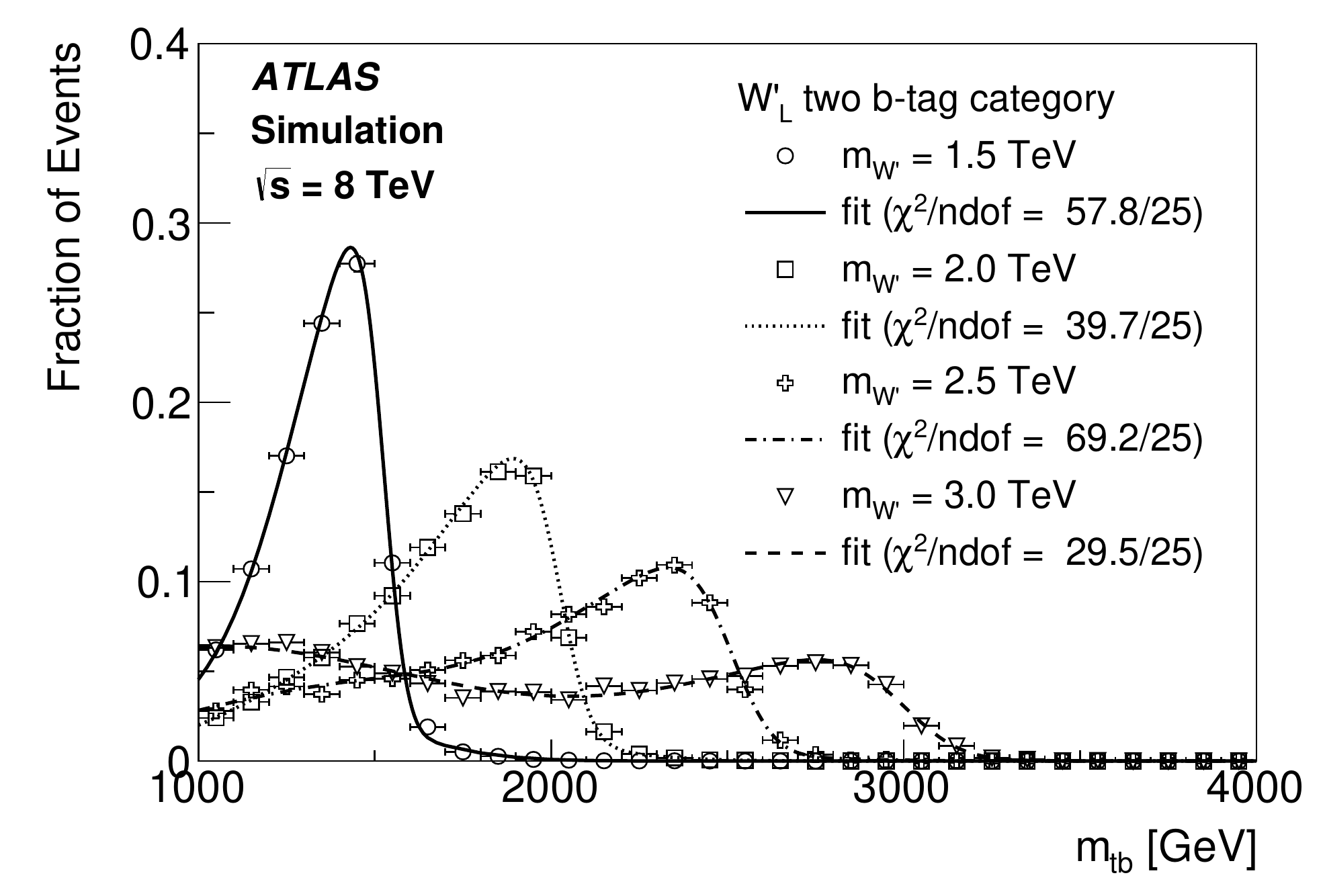}
  \caption{Parametric fits to the $\wp\rightarrow tb \rightarrow qqbb$ signal distributions in the two $b$-tag category for \wpl\ masses between 1.5 and 3~\tev\  overlaid
    to the corresponding MC distributions.
    While unbinned fits are performed, the events from MC simulation are shown as a histogram for presentational purposes.
  }
  \label{fig:shapesignal}
\end{figure}

The reconstructed $m_{tb}$ spectrum for the \wp\ signal is parametrised using the sum of a skew-normal~\cite{skewnormal} and a Gaussian function.
The skew-normal accounts for the asymmetric shape of the resonant \wp\ signal and is the product of a Gaussian and a Gaussian error function.
Non-resonant off-shell \wp\ production is accounted for with the additional Gaussian distribution.
This allows the signal shape to be fully parametrised.
In order to search for a \wp\ signal over the full mass range, the signal shape parameters, as well as the signal acceptance and the expected cross sections are interpolated between the generated mass points.
Figure~\ref{fig:shapesignal} shows the parametric fits to the \wp\ signal distributions in the two $b$-tag category for \wpl\ masses between 1.5 and 3~\tev\ overlaid to the corresponding MC distributions.

\begin{table*}[t]
\begin{center}
  \caption{\label{tab:syst2tev} Systematic uncertainties on the event yields of a 2~\tev\ \wp\ boson in both categories in percent and on the
    background modelling in numbers of expected \wp\ boson events.}
 \begin{tabular}{lcccc}
   \hline
   \multicolumn{1}{l}{Systematic Uncertainties [\%]}& \multicolumn{2}{c}{\wpl} & \multicolumn{2}{c}{\wpr}\\
   & One $b$-tag & Two $b$-tag & One $b$-tag & Two $b$-tag\\
   \hline
$b$-tagging                &  +13, -20  &  +45, -37   &   +15, -21   &  +40, -34   \\
W' Top-tagging             &  $\pm$13   &  $\pm$ 10   &   $\pm$ 11   &  $\pm$ 9    \\
Jet Energy Scale           &  $\pm$1.3  &  $\pm$ 1.9  &   $\pm$ 0.8  &  $\pm$ 1.9  \\
Jet Energy Resolution      &  $\;<$0.1 &  $\pm$ 0.2  &   $\pm$ 0.1  &  $\pm$ 0.5  \\
   Theoretical             & \multicolumn{2}{c}{$\pm 10$} & \multicolumn{2}{c}{+8, -10} \\
   Luminosity              & \multicolumn{4}{c}{$\pm$2.8} \\
   \hline
   Background Modelling     & $\pm$44 events & $\pm$28 events & $\pm$45 events & $\pm$24 events \\
   \hline
   \end{tabular}
 \end{center}
\end{table*}

Additional contributions from $\wp \rightarrow tb \rightarrow \ell\nu b b$ are between 3.5 and 10\%.
The leptonic contribution is taken into account by fitting the reconstructed $m_{tb}$
distribution with a double-Gaussian function and interpolating between generated masses analogously to the treatment of the $\wp \rightarrow tb \rightarrow qqbb$ signal.
Large-$R$ jets can falsely be top-tagged in events with leptonic top-quark decays due to several effects including hard gluon radiation, calorimeter activity from
non-identified electrons and hadronically decaying tau leptons.

\begin{figure*}[p]
  \centering
  \includegraphics[width=0.4\textwidth]{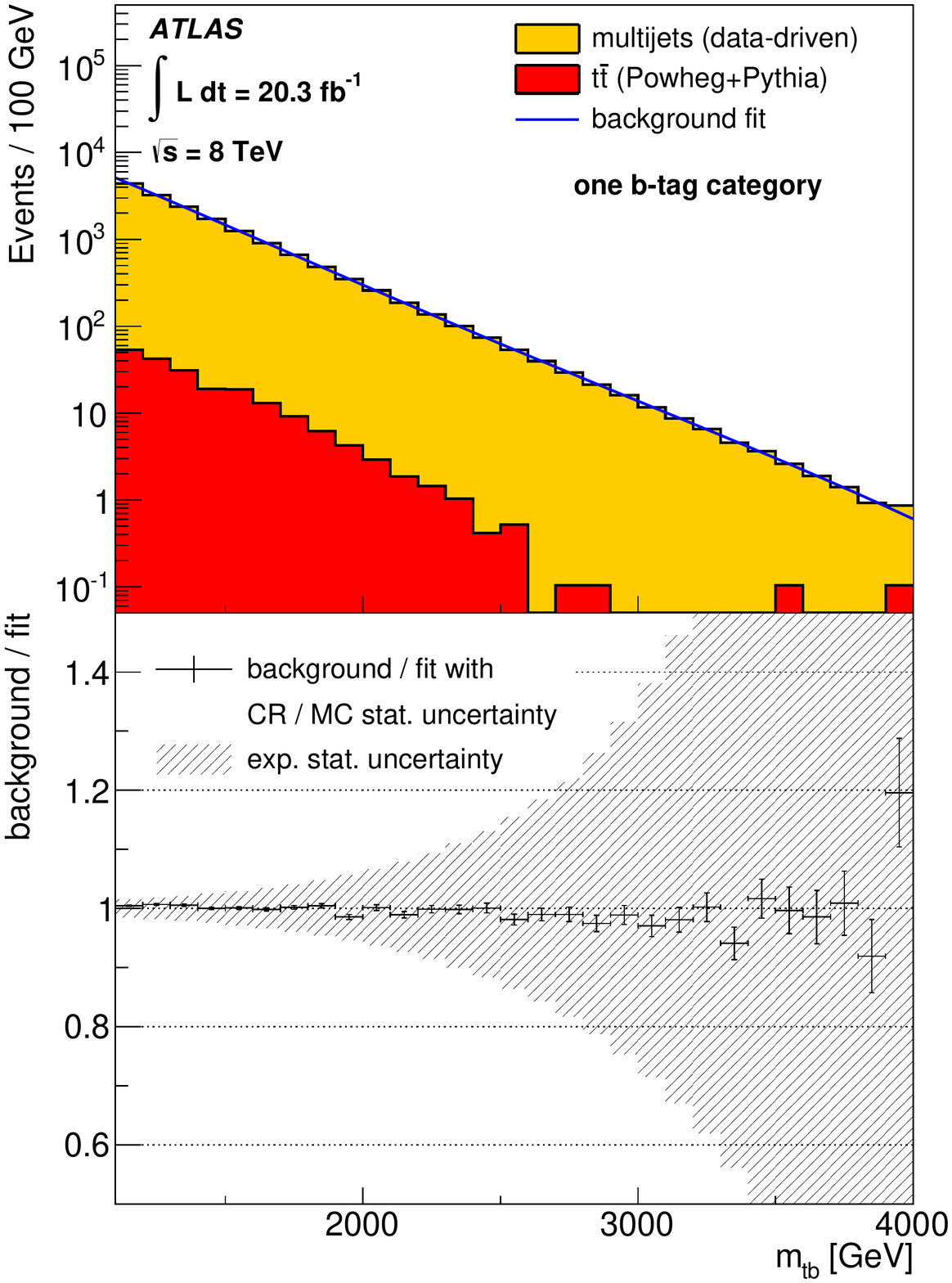}
  \includegraphics[width=0.4\textwidth]{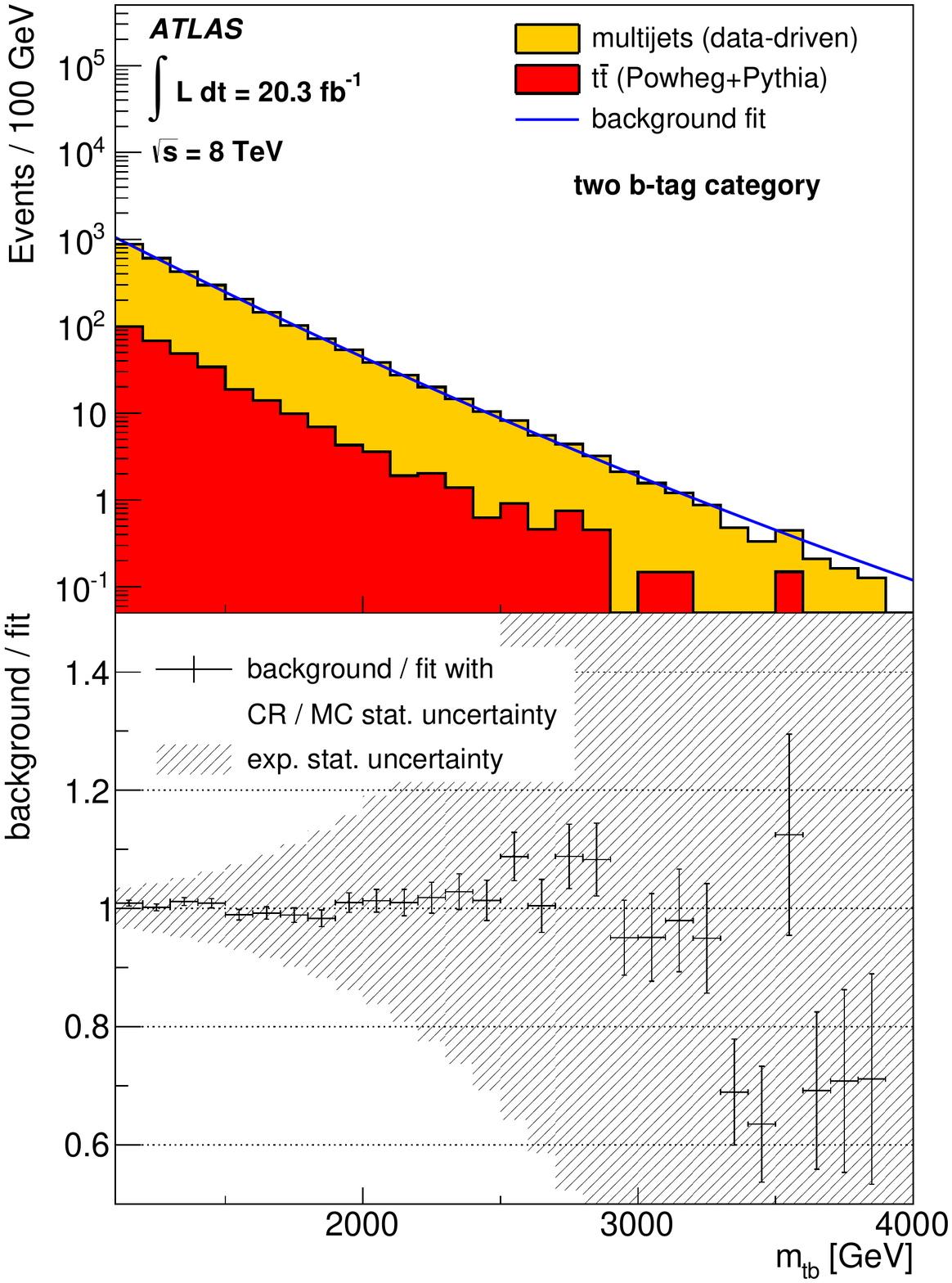}
  \caption{Fit to the $m_{tb}$ distribution from the initial background estimate in the one $b$-tag category (left) and in the two $b$-tag category (right).
    The ratio background/fit is also shown, where the uncertainties are the control-region and MC statistical uncertainties and the grey shaded band shows the
    statistical uncertainty in each bin as expected for \lumi.
  }
  \label{fig:ABCDbkgfit_differentfunctions_2}
  \vspace{1cm}
  \centering
  \includegraphics[width=0.4\textwidth]{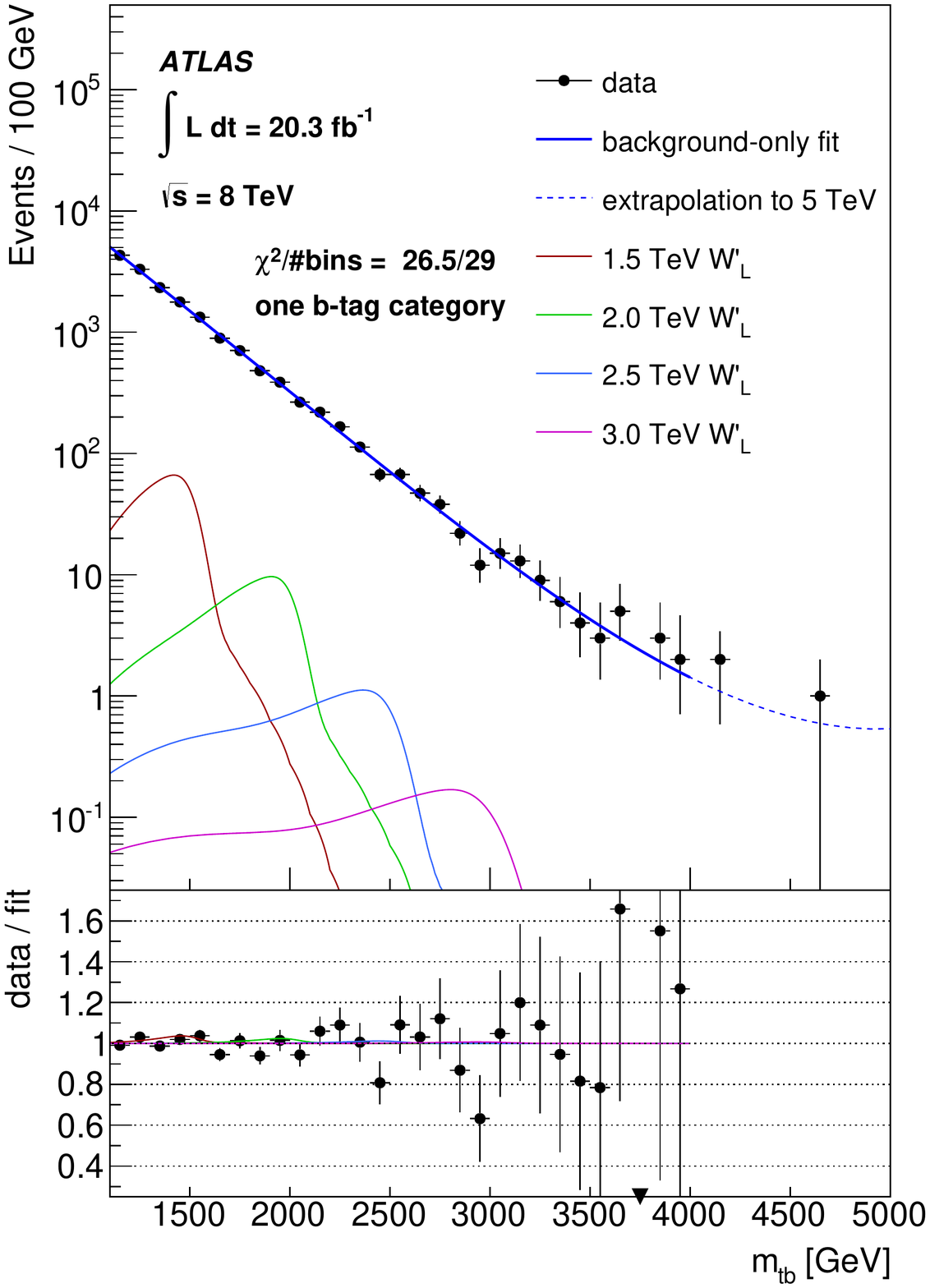}
  \includegraphics[width=0.4\textwidth]{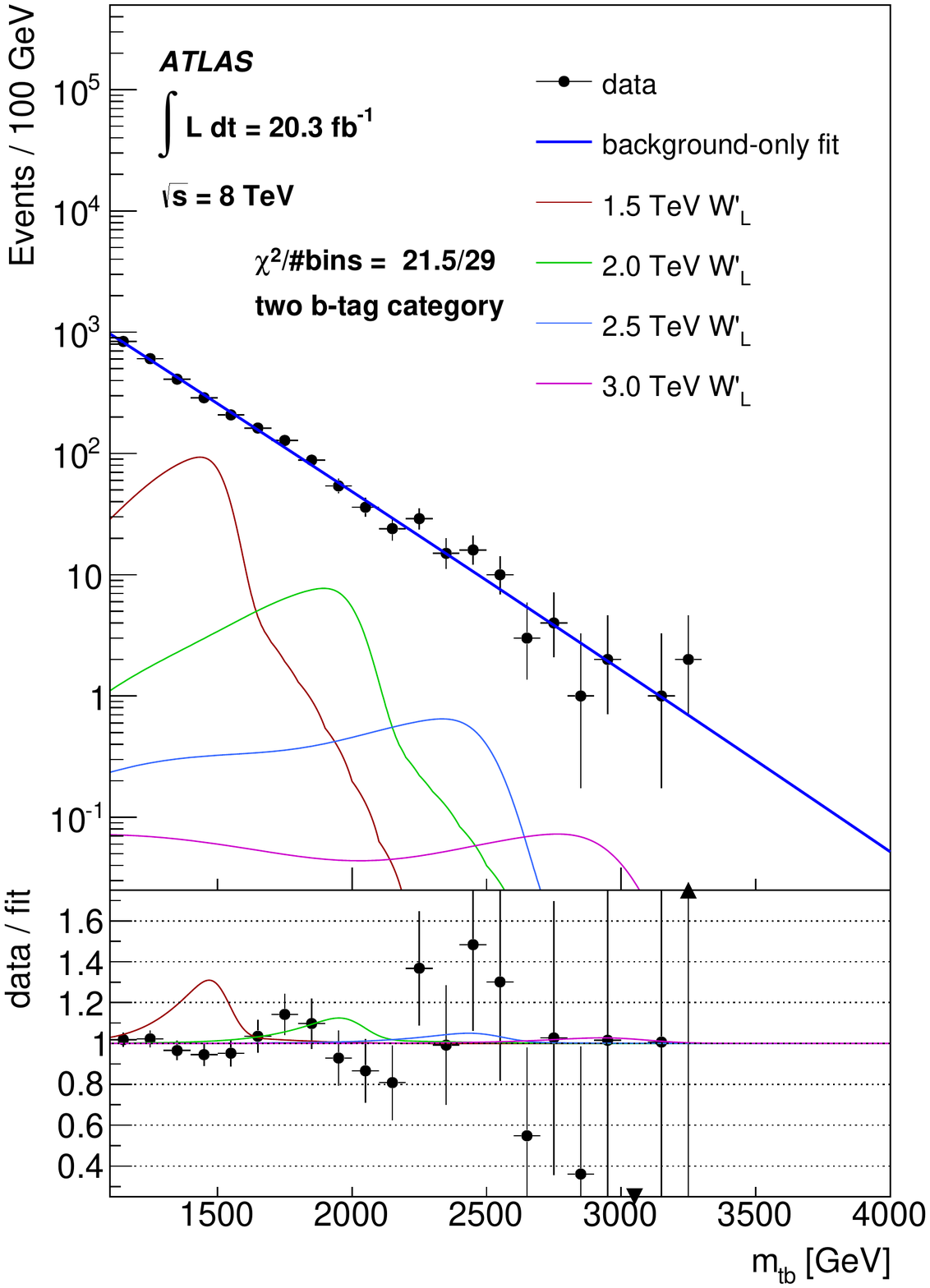}
  \caption{$m_{tb}$ distributions in data in the one $b$-tag (left) and the two $b$-tag category (right).
    Background-only fits are shown, and the bottom plots show the ratio of the data and the fit.
    The left plot shows an extrapolation of the background fit into the region 4~--~5~\tev. The ratio plot, however does not show the three data points in this range,
    because they are beyond the range considered for this analysis.
    Potential \wpl\ signal shapes in the hadronic top-quark decay channel with $g'=\gSM$
    are also overlaid for resonance masses of 1.5, 2.0, 2.5 and 3.0~\tev.
  }
  \label{fig:backgroundshape}
\end{figure*}

The sum of all backgrounds is fitted with an analytic function. For each of the two categories, a function
is chosen among several tested functions following a procedure based on the $m_{tb}$ distribution obtained from the initial background estimate
(Section~\ref{sec:hadbackground}).
For each function under study the corresponding fake signal bias is quantified by fitting the $m_{tb}$ distribution from the initial background estimate
with a background-plus-signal model.
The maximal extracted fake signal observed over the full range of \wp\ masses is chosen as systematic uncertainty on the background modelling (Section~\ref{sec:systuncert}).
The function with the least number of free parameters is chosen out of all tested functions giving similarly small bias.
This procedure yields an exponential function with a polynomial
of order $n$ as its argument:
\mbox{$\exp\left( \sum_{k=1}^{n} c_k m_{tb}^k \right)$}, with $n = 4$ ($n = 2$) in the one (two) $b$-tag category.
Figure~\ref{fig:ABCDbkgfit_differentfunctions_2} shows background-only fits to the initial background estimate
in the one and two $b$-tag categories using the chosen functions.
The ratio of the distribution and the fit is also shown, where the uncertainties are the control-region and MC statistical uncertainties and the grey shaded band
shows the statistical uncertainty in each bin as expected for \lumi.
Deviations from 1 in the ratio plot are much smaller than the expected statistical uncertainty and the chosen background functions are hence shown to be flexible
enough to describe the background distribution in the signal region.

\section{Systematic Uncertainties}
\label{sec:systuncert}

Systematic uncertainties may change the acceptance and shape of the potential \wp\ signal, and are included as nuisance parameters in the likelihood function.
Table~\ref{tab:syst2tev} shows the impact of the systematic uncertainties on the event yield of a 2~\tev\ \wp\ boson in the one and two $b$-tag categories.
The largest sources of uncertainty come from the uncertainties associated with $b$-tagging, top-tagging and background modelling.

Uncertainties on the $b$-tagging efficiency and mistag rates are estimated from data using $\ttbar$ di-lepton decays~\cite{ATLAS-CONF-2014-046,ATLAS-CONF-2014-004}.
The $b$-tagging (mistagging) uncertainties are increased for high \pt\ and reach up to 34\% (60\%) per jet.
Uncertainties on the \wp\ top-tagger performance are evaluated based on the data-MC agreement as shown in Refs.~\cite{groomedCONF2011,groomedJetPileup2011}.
They are derived comparing the ratio of each variable from jets built from calorimeter clusters and the corresponding jet built from tracks in the ID.
The observed differences between data and MC are taken as variations on the substructure variables and are translated into an uncertainty on the efficiency of the \wp\ top-tagger.
Within the kinematic reach, it has been shown with $\ttbar$ events in the single-lepton channel that these uncertainties cover any possible disagreement
between the efficiency observed in data and MC simulations.
The jet energy scale (JES) uncertainty~\cite{ATLAS-CONF-2013-004}
depends on the \pt\ and $\eta$\ of the reconstructed jet and includes the uncertainty on the $b$-jet energy scale. The JES of the two jet types are assumed to be correlated.
The impact of the jet energy resolution uncertainty is evaluated by smearing the jet energy in the simulation to increase the nominal resolution.
 The uncertainty on the integrated luminosity is 2.8\% as derived from beam-separated scans~\cite{lumiref}. 
Theoretical uncertainties are included by evaluating the change in the expected number of signal events. 
The deviations from varying the {\sc CTEQ6L1} PDF eigenvectors are summed in quadrature with the uncertainty from $\alpha_s$, the renormalisation scale, and the change in acceptance at LO and NLO.
In addition, the uncertainty on the beam energy~\cite{beamerror} is included.

Uncertainties due to background mismodelling are quantified as discussed in Section~\ref{sec:analsys}.
This uncertainty amounts to 28 (24) events in the two $b$-tag category and 44 (45) events in the one $b$-tag category for the \wpl\ (\wpr) model.

\begin{figure*}[p]
  \centering
  \includegraphics[width=0.48\textwidth]{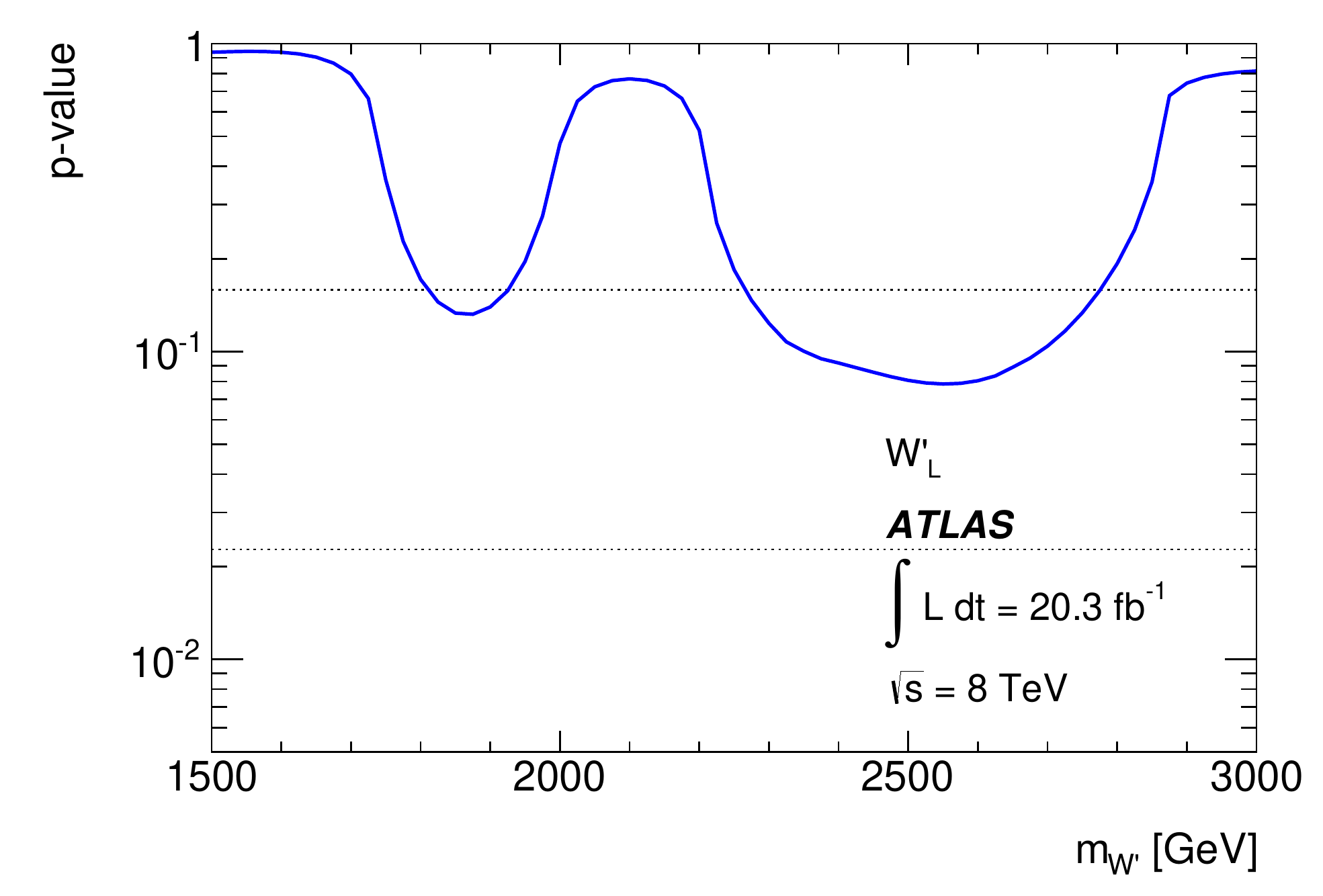}
  \includegraphics[width=0.48\textwidth]{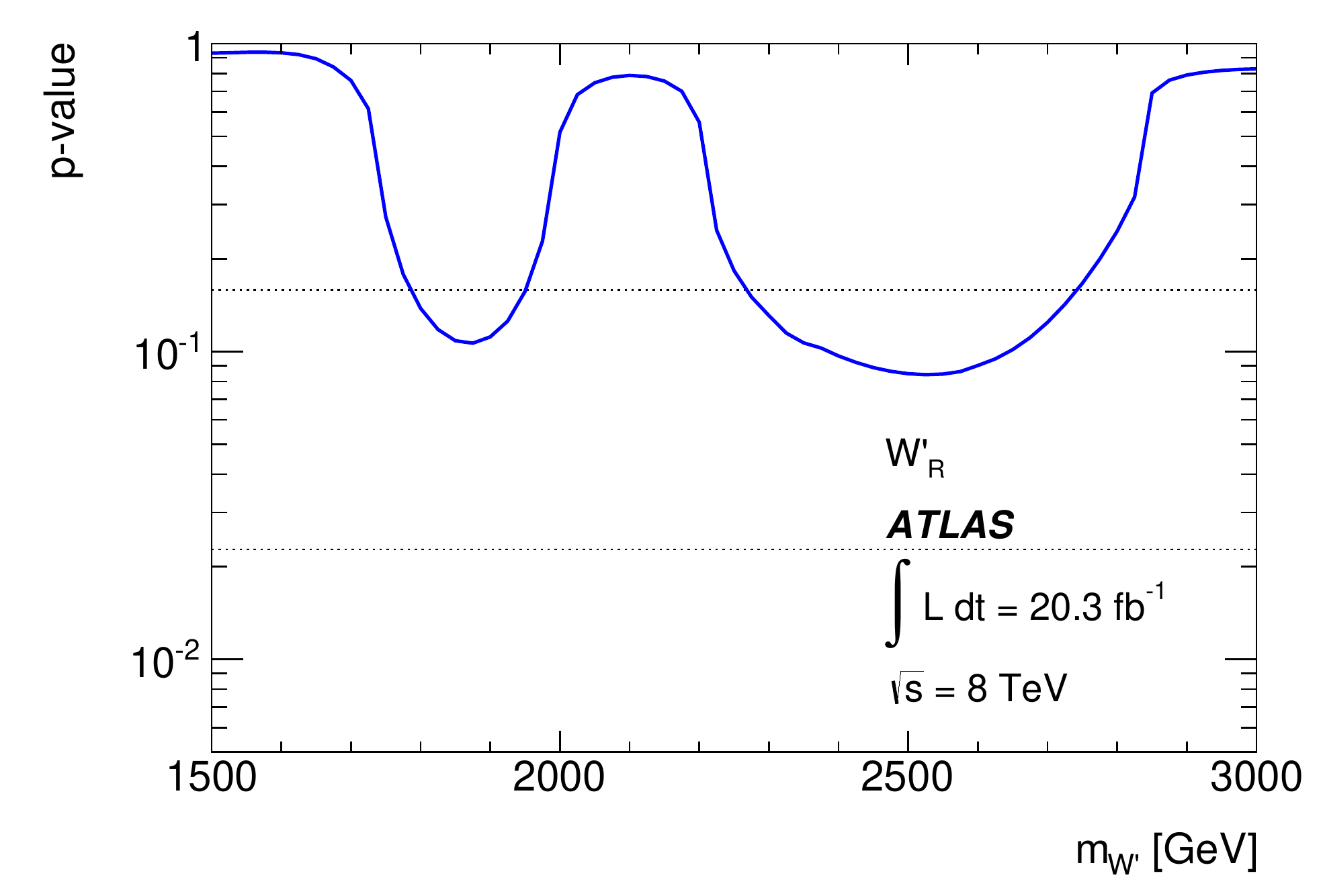}
  \caption{Observed $p$-values for a left-handed (left) and right-handed (right) \wp\ model as a function of the \wp\ mass.}
  \label{fig:pvalues_observed}
  \vspace{1cm}
  \centering
  \includegraphics[width=0.48\textwidth]{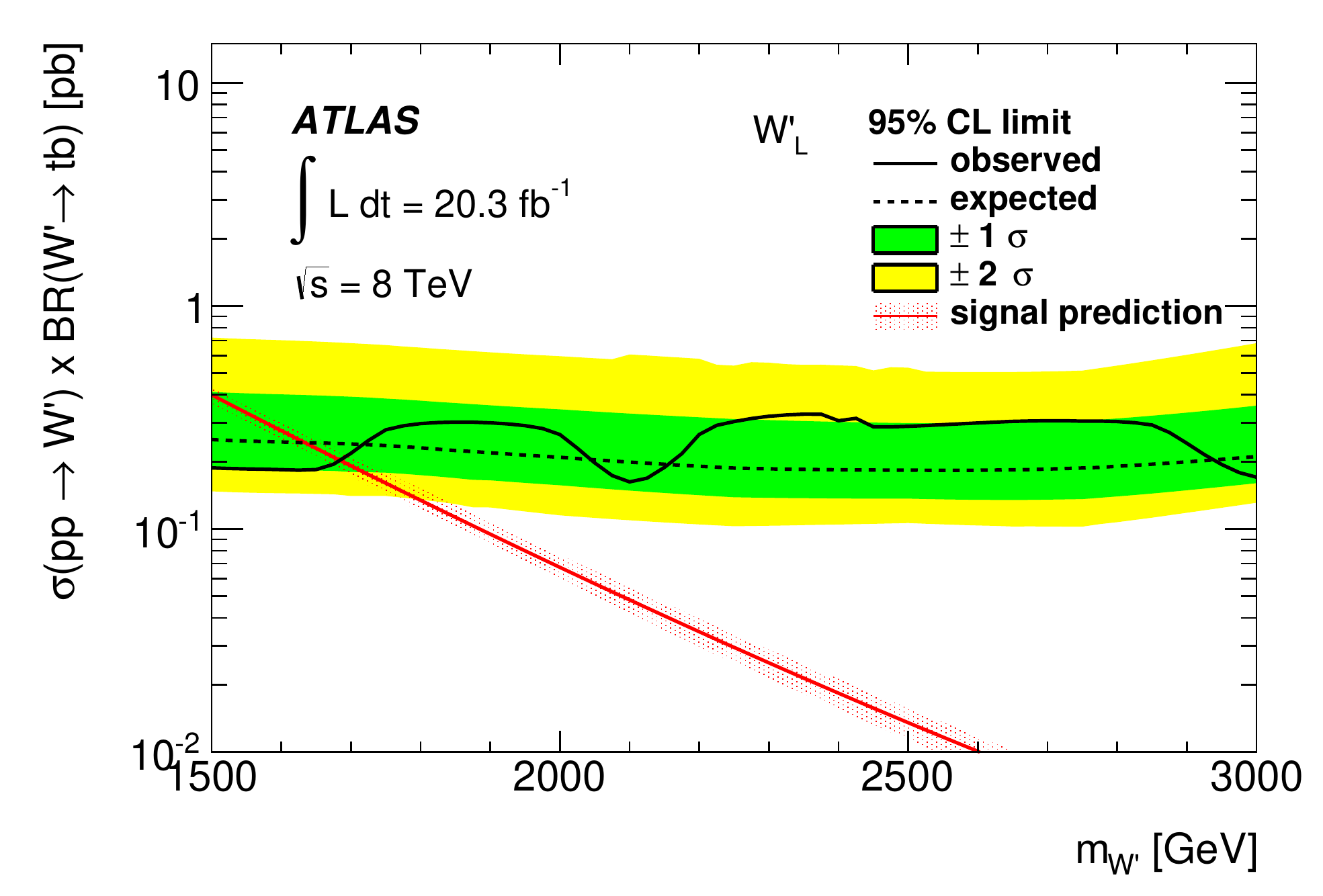}
  \includegraphics[width=0.48\textwidth]{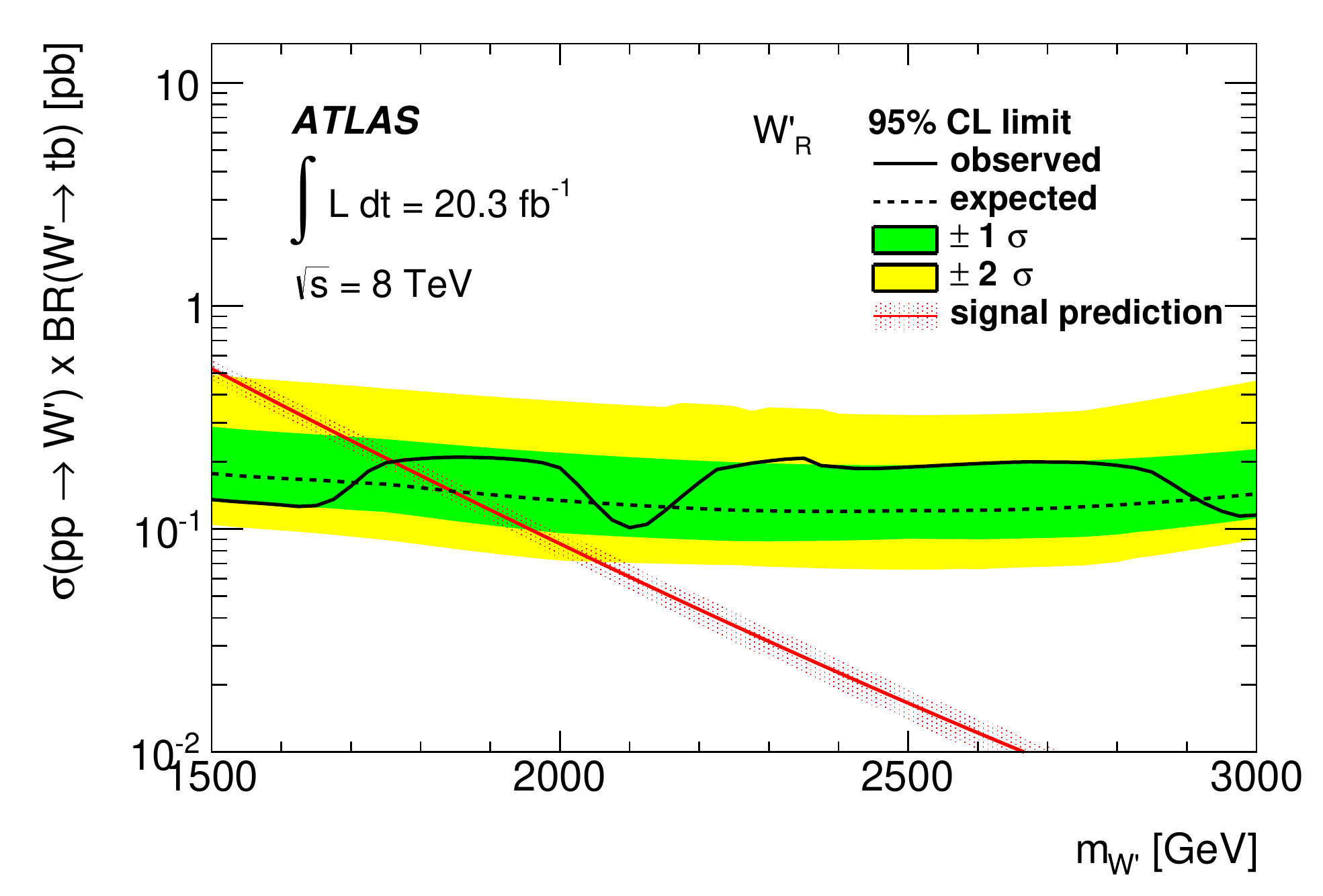}
  \caption{Limits at 95\% CL on the cross section times branching ratio to $tb$ for the left-handed (left) and for the right-handed (right) \wp\ model.
  The expected cross section for \wp\ production with $g'=\gSM$ is also shown.}
  \label{fig:limits_observed}
  \vspace{1cm}
  \centering
  \includegraphics[width=0.48\textwidth]{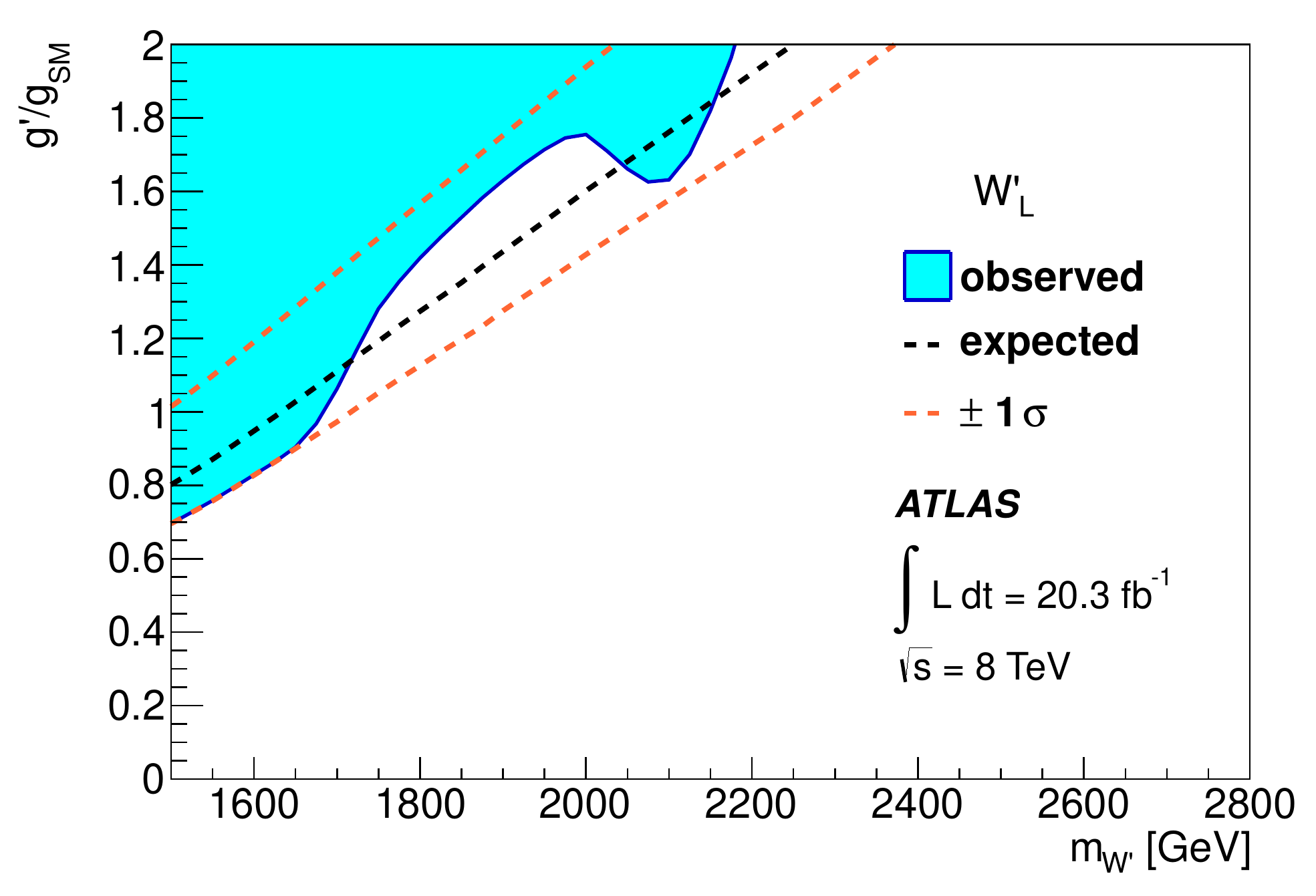}
  \includegraphics[width=0.48\textwidth]{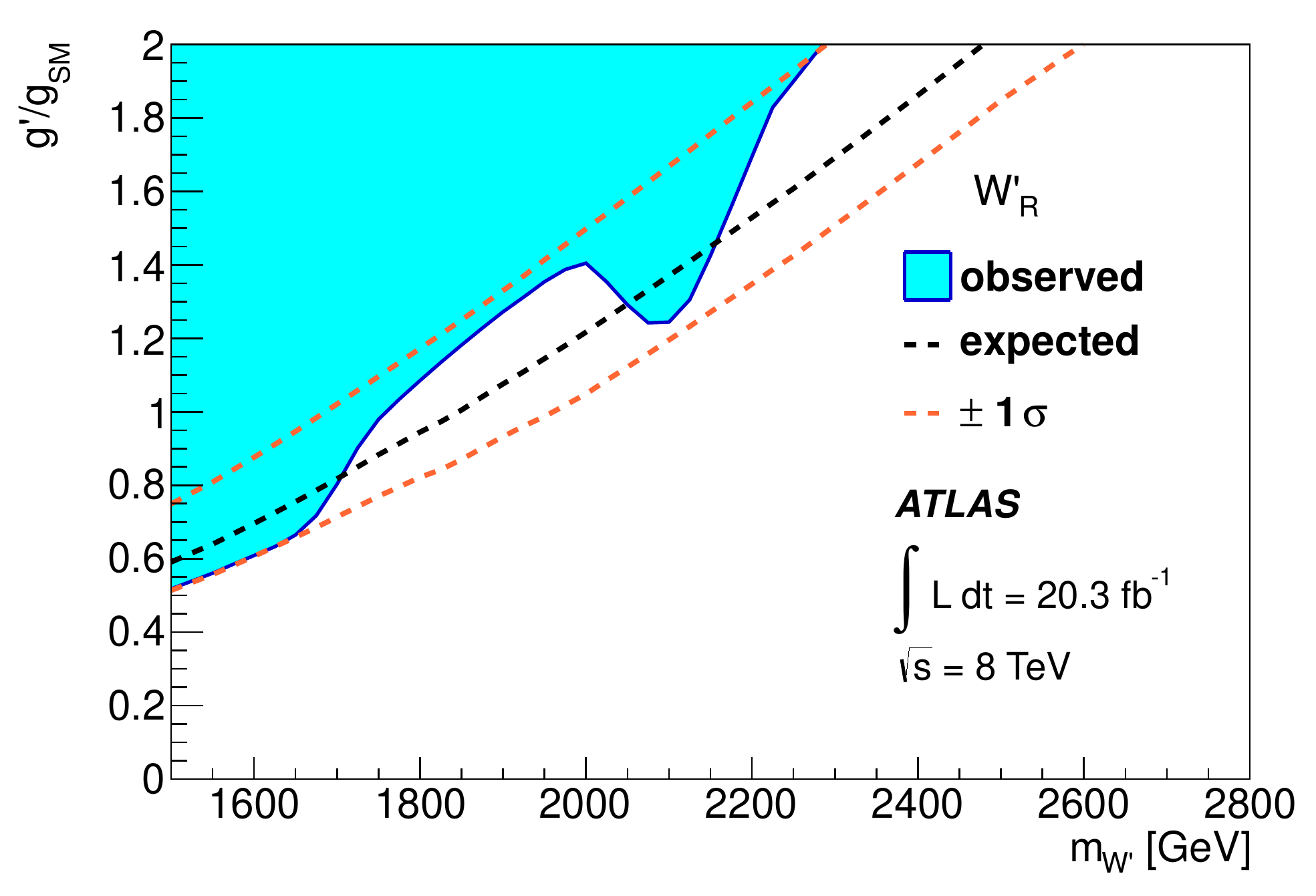}
  \caption{Observed and expected 95\% CL limits on the ratio of coupling $g'_L/\gSM$ ($g'_R/\gSM$) of the \wpl\ (\wpr) model as a function of the \wp\ mass. }
  \label{fig:limit1d_2d}
\end{figure*}

\section{Results}
\label{sec:result}

Figure~\ref{fig:backgroundshape} shows the observed $m_{tb}$ spectra in the two categories.
The highest mass event in the two (one) $b$-tag category is at 3.25~\tev\ (4.68~\tev).
A background-only fit to the spectra is also shown and good agreement is observed between the fit and the data. 

Figure~\ref{fig:pvalues_observed} shows the observed $p$-values for background plus left-handed or right-handed \wp\ model as a function of the \wp\ mass allowing the background parameters to float.
The $p$-value from both categories combined is shown taking into account all systematic uncertainties. The maximum local significance is 1.4$\sigma$.
Hence, no significant excess over the background-only hypothesis is observed and 95\% CL limits are derived on the cross section times branching
ratio to $tb$ as shown in Figure~\ref{fig:limits_observed}. 
The observed limits agree with the expected limits roughly within 1$\sigma$ over the whole range for both \wp\ models and
range from 0.16~pb to 0.33~pb for left-handed \wp\ bosons and from 0.10~pb to 0.21~pb for \wp\ bosons with purely right-handed couplings. The relative mass independence of the expected sensitivity and observed limits is a unique feature of this search arising from the flat signal efficiency when combining the one and two $b$-tag categories.

For $g' = \gSM$, the limits on the cross section times branching ratio translate to observed (expected) limits on the mass to be above
\limitleft\ (\explimitleft) and \limitright\ (\explimitright) in the left- and right-handed models, respectively.
The observed cross-section limits are also interpreted as limits on other values of the couplings $g'_{L/R}$.
For $g'_{L/R}/\gSM < 2$, the reconstructed $m_{tb}$ distributions are dominated by the experimental width.
The results obtained for $g'_{L/R} = \gSM$ can hence be interpreted as limits in the $g'_L$-\wpl\ mass
($g'_R$-\wpr\ mass) plane, making use of the approximately quadratic dependence of the \wp\ production cross section on $g'$.
The observed and expected limits on the ratio of couplings $g'_L/\gSM$ ($g'_R/\gSM$) of the \wpl\ (\wpr) model as a function of
\wp\ mass are shown in Figure~\ref{fig:limit1d_2d} and amount to $g' < 0.70$ ($g' < 0.55$) for a 1.5~\tev\ \wpl\ (\wpr) and
to $g' < 2$ for a 2.18 (2.29)~\tev\ \wpl\ (\wpr) boson.

\section{Summary and Conclusion}
\label{sec:summary}
A search for $\wp \rightarrow tb \rightarrow qqbb$ was presented using \lumi\ of 8~\tev\ proton-proton collisions data taken with the ATLAS detector. 
The analysis makes use of jet substructure tagging optimised to select large-$R$ jets coming from hadronically decaying top quarks and $b$-tagging of small-$R$ jets. 
The observed $m_{tb}$ spectrum from data is consistent with the background-only prediction and
exclusion limits at 95\% CL are set on the \wp\ boson production cross section times branching ratio to $tb$.
The use of novel jet substructure techniques allows cross-section limits to be set at high \wp\ masses, which are similar to the limits at lower masses and range
from 0.16~pb to 0.33~pb for left-handed \wp\ bosons,
and from 0.10~pb to 0.21~pb for \wp\ bosons with purely right-handed couplings.
In addition, limits are set at 95\% CL on the \wp-boson effective couplings as a function of the \wp\ mass.

\section*{Acknowledgments}

We thank CERN for the very successful operation of the LHC, as well as the
support staff from our institutions without whom ATLAS could not be
operated efficiently.

We acknowledge the support of ANPCyT, Argentina; YerPhI, Armenia; ARC,
Australia; BMWFW and FWF, Austria; ANAS, Azerbaijan; SSTC, Belarus; CNPq and FAPESP,
Brazil; NSERC, NRC and CFI, Canada; CERN; CONICYT, Chile; CAS, MOST and NSFC,
China; COLCIENCIAS, Colombia; MSMT CR, MPO CR and VSC CR, Czech Republic;
DNRF, DNSRC and Lundbeck Foundation, Denmark; EPLANET, ERC and NSRF, European Union;
IN2P3-CNRS, CEA-DSM/IRFU, France; GNSF, Georgia; BMBF, DFG, HGF, MPG and AvH
Foundation, Germany; GSRT and NSRF, Greece; RGC, Hong Kong SAR, China; ISF, MINERVA, GIF, I-CORE and Benoziyo Center, Israel; INFN, Italy; MEXT and JSPS, Japan; CNRST, Morocco; FOM and NWO, Netherlands; BRF and RCN, Norway; MNiSW and NCN, Poland; GRICES and FCT, Portugal; MNE/IFA, Romania; MES of Russia and ROSATOM, Russian Federation; JINR; MSTD,
Serbia; MSSR, Slovakia; ARRS and MIZ\v{S}, Slovenia; DST/NRF, South Africa;
MINECO, Spain; SRC and Wallenberg Foundation, Sweden; SER, SNSF and Cantons of
Bern and Geneva, Switzerland; NSC, Taiwan; TAEK, Turkey; STFC, the Royal
Society and Leverhulme Trust, United Kingdom; DOE and NSF, United States of
America.

The crucial computing support from all WLCG partners is acknowledged
gratefully, in particular from CERN and the ATLAS Tier-1 facilities at
TRIUMF (Canada), NDGF (Denmark, Norway, Sweden), CC-IN2P3 (France),
KIT/GridKA (Germany), INFN-CNAF (Italy), NL-T1 (Netherlands), PIC (Spain),
ASGC (Taiwan), RAL (UK) and BNL (USA) and in the Tier-2 facilities
worldwide.

\bibliographystyle{atlasnote}
\bibliography{ATLAS,wprime,MC}

\clearpage
\newpage

\onecolumn
\clearpage 
\begin{flushleft}
{\Large The ATLAS Collaboration}

\bigskip

G.~Aad$^{\rm 84}$,
B.~Abbott$^{\rm 112}$,
J.~Abdallah$^{\rm 152}$,
S.~Abdel~Khalek$^{\rm 116}$,
O.~Abdinov$^{\rm 11}$,
R.~Aben$^{\rm 106}$,
B.~Abi$^{\rm 113}$,
M.~Abolins$^{\rm 89}$,
O.S.~AbouZeid$^{\rm 159}$,
H.~Abramowicz$^{\rm 154}$
% ...
\bigskip
\\
% ...
$^{11}$ Institute of Physics, Azerbaijan Academy of Sciences, Baku, Azerbaijan\\
% ...
$^{84}$ CPPM, Aix-Marseille Universit{\'e} and CNRS/IN2P3, Marseille, France\\
% ...
$^{89}$ Department of Physics and Astronomy, Michigan State University, East Lansing MI, United States of America\\
% ...
$^{106}$ Nikhef National Institute for Subatomic Physics and University of Amsterdam, Amsterdam, Netherlands\\
% ...
$^{112}$ Homer L. Dodge Department of Physics and Astronomy, University of Oklahoma, Norman OK, United States of America\\
$^{113}$ Department of Physics, Oklahoma State University, Stillwater OK, United States of America\\
% ...
$^{116}$ LAL, Universit{\'e} Paris-Sud and CNRS/IN2P3, Orsay, France\\
% ...
$^{152}$ Institute of Physics, Academia Sinica, Taipei, Taiwan\\
% ...
$^{154}$ Raymond and Beverly Sackler School of Physics and Astronomy, Tel Aviv University, Tel Aviv, Israel\\
% ...
$^{159}$ Department of Physics, University of Toronto, Toronto ON, Canada\\
% ...
\end{flushleft}

\end{document}